\documentclass[%
 reprint,
showpacs,
 amsmath,amssymb,
 aps,
]{revtex4-1}

\usepackage{graphicx}
\usepackage{dcolumn}
\usepackage{bm}
\usepackage{amsmath}
\usepackage{amssymb}
\usepackage{braket}
\usepackage{physics}
\usepackage{amsthm}
\usepackage{natbib}
\usepackage{mathtools}
\usepackage{diagbox}
\usepackage[utf8]{inputenc}
\usepackage[english]{babel}
\DeclarePairedDelimiter{\ceil}{\lceil}{\rceil}

\begin{document}


\title{Feynman-path type simulation using stabilizer projector decomposition of unitaries}

\author{Yifei Huang}\thanks{yifei.huang@tufts.edu} 

\author{Peter Love}\thanks{Corresponding author: peter.love@tufts.edu}
\affiliation{Department of Physics and Astronomy, Tufts University.}

\begin{abstract}

We propose a classical simulation method for quantum circuits based on decomposing unitary gates into a sum of stabilizer projectors. By only decomposing the non-Clifford gates, we take advantage of the Gottesman-Knill theorem and build a bridge between stabilizer-based simulation and Feynman-path type simulation. We give two variants of this method: stabilizer-based path integral recursion~(SPIR) and stabilizer projector contraction~(SPC). We also analyse further advantages and disadvantages of our method compared to the Bravyi-Gosset algorithm and recursive Feynman path-integral algorithms. We construct a parameterized circuit ensemble and identify the parameter regime in this ensemble where our method offers superior performance. We also estimate the time cost for simulating quantum supremacy experiments with our method and motivate potential improvements of the method.  
\end{abstract}

\maketitle

\section{Introduction}

Noisy Intermediate-Scale Quantum~(NISQ) devices are quantum systems where the quality of the qubits and gates are not sufficient to enable fault-tolerant quantum computation but whose output probability distributions are still hard for classical devices to sample from~\cite{preskill2018quantum}. Improvements in the performance of NISQ devices are stimulating the development of methods to simulate them on classical computers~\cite{boixo2017simulation,markov2018quantum, chen201864,chen2018classical,guo2019general,villalonga2020establishing,zhou2020limits,huang2020classical,barak2020spoofing}. Specifically, when running quantum algorithms to demonstrate quantum supremacy~\cite{preskill2012quantum,boixo2018characterizing,arute2019quantum}, one needs a classical machine to verify the outputs of the NISQ device. Improved classical simulation can improve the benchmarking of NISQ devices, as well as potentially challenging current supremacy results~\cite{zhou2020limits,huang2020classical}.

Simulation of quantum mechanics in general, and quantum computation in particular, is believed to be hard. However, proofs of the difficulty of the simulation of quantum mechanics remain elusive. For example, efficient classical simulation of Shor's factoring algorithm would imply that factoring is classically easy. The widespread disbelief in the classical tractability of factoring is the basis for the use of the RSA cryptosystem. Nevertheless, the classical difficulty of factoring remains unproven. 

The exponentially rapid growth of Hilbert space with number of qubits is frequently invoked to underpin arguments concerning the difficulty of classical simulation of quantum systems. In fact the simulation of quantum systems only requires polynomial memory~\cite{nielsen2011quantum}. However, methods that use polynomial memory do so by computing the amplitude of one path at a time in a path integral representation. Hence, the claim of exponential cost of classical simulation rests on the belief that it is necessary to compute the amplitude of an exponentially large number of paths. 

The difficulty in finding an efficient classical simulation algorithm for quantum mechanics arises from the nature of the wavefunction. As Heisenberg said, the wavefunction is partly a thing, and partly our knowledge of a thing~\cite{heisenberg1959physics}. It remains unclear to what extent the wavefunction is ontological in nature - meaning it represents a thing, or epistemological in nature, meaning it represents our knowledge of a thing~\cite{pusey2012reality,leifer2014quantum,fuchs2014introduction}. 

If the wavefunction were ontological - meaning that simulation of the entire wavefunction was inescapable - the difficulty of simulation of quantum mechanics would be clear. If the wavefunction is epistemic one could treat simulation of quantum mechanics as one does probabilistic simulations of classical systems - by sampling and other techniques that might yield an efficient scheme in particular cases. 

These two points of view on the wavefunction lead to two types of classical simulation: strong and weak, In strong simulation, the simulator is given a string $x$ and it outputs the exact output probability $p(x)=\abs{\bra{x}U\ket{0^n}}^2$ of observing $x$, or it approximates the probabilities to a multiplicative error:
\begin{equation}
(1-\epsilon)\hat{p}(x)\leq p(x) \leq (1+\epsilon)\hat{p}(x), \label{Multi_error}  
\end{equation}
where $\hat{p}(x)$ is the estimated probability. 

In weak simulation, the outputs of the simulator are samples from $p(x)$ instead of probabilities, see Appendix \ref{notion} for more discussion of the notion of weak simulation. In this paper, we only consider the notion of exact strong simulation.

The direct simulation of $n$-qubit quantum evolution stores the whole wave function and evolves it by applying unitary operators. Direct simulation requires memory $O(2^n)$ and time $O(m2^n)$, where $m$ is the number of gates in the circuit. We refer to this as the Schrodinger approach in accordance with~\cite{arute2019quantum}. In the following, we will also briefly review the recursive Feynman path-integral simulation algorithm and the Bravyi-Gosset algorithm for simulating quantum circuits, as they will be relevant for our later discussion.

A circuit has depth $d$ when the circuit can be divided into $d$ layers where each layer has gates acting on disjoint sets of qubits. The width $w_i$ of each layer will be the number of qubits that are acted on by gates in each layer. If the width for all layers is $O(w)$, then the number of gates $m$ will be $O(dw)$. Even for shallow NISQ devices, we know that the depth of a quantum circuit at least needs to be more than 2 to be hard for classical computers to simulate~\cite{terhal2002adaptive}. Assuming the width for all layers is $O(n)$, then the depth-3 circuit that is hard for classical simulation has $m\approx O(3n)$ gates. Therefore, we have $m>n$ in general. In this paper, we also define the non-Clifford depth $d_{nc}$, which we will use in Section \ref{alg}. We divide the circuit into Clifford layers and non-Clifford layers and the number of non-Clifford layers is the non-Clifford depth $d_{nc}$ in that context.

\subsection{Recursive Feynman path-integral algorithm}\label{Recur_Path}

The Feynman approach calculates the amplitude of obtaining outcome of string $x$ using the path-integral representation of a circuit:
\begin{equation}
\begin{split}
&\bra{x}U_mU_{m-1}...U_1\ket{0^n}\\&=\sum_{j_1,...,j_{m-1}} \bra{x}U_m\ket{j_{m-1}}\bra{j_{m-1}}U_{m-1}\ket{j_{m-2}}...\bra{j_1}U_1\ket{0^n}
\end{split}
\label{path_sim}   
\end{equation}
This method requires memory $O(m+n)$, because after calculating each amplitude, one only needs to add the final result to the sum. However, $O(4^m)$ operations are required to process all possible paths. Because $m$ is almost always bigger than $n$, the number of operations $O(4^m)$ here is larger than that of the Schrodinger approach.

To reduce the time cost for the Feynman approach, one can notice that there are a large number of repeated calculations in eq. (\ref{path_sim}) if we calculate the terms in the sum path by path. For example, if two of the paths are the same for the first $m-2$ steps, but only differs at the last step. In eq. (\ref{path_sim}), these two paths will give two terms that only differ on the first factor of the term, $\bra{x}U_m\ket{j_{m-1}}$. However, naive evaluation of eq.~(\ref{path_sim}) results in redundant computation of the first $m-1$ factors.

As discussed in \cite{aaronson2017complexity}, one can avoid this repeated calculation as follows. By first slicing a circuit into two sub-circuits first, $C_1$ and $C_2$:
\begin{equation}
\bra{x}C\ket{0^n}=\sum_{y\in \{0,1\}^n}\bra{x}C_2\ket{y}\bra{y}C_1\ket{0^n}.  
\end{equation}
one can obtain a recursion relation for the time cost for calculating the whole sum from the results of the two sub-circuits:
\begin{equation}
T(d)=2^{n+1}T(d/2)   
\end{equation}
assuming the depth of the whole circuit is $d$ and the depth of both $C_1$ and $C_2$ is $d/2$. Following this relation, one can recursively divide the two sub-circuits further until getting down to single-layer circuits. In this way, one can calculate the sum more efficiently than the Feynman approach, without sacrificing much of the space cost advantages of the Feynman approach~\cite{aaronson2017complexity}. In fact, we only need $O(\log d)$ steps to reach the leaf level~(single layer) with recursion, and at each step we need a $n$-bit string $y$ to label the term we are trying to compute. Therefore, we need $O(n\log d)$ space to recursively return a single term to the whole summation. Meanwhile, one can see the total time cost is brought down to $O(n2^{n\log d})$ by solving the recursion relation above. We will use the idea of this algorithm later when we apply this recursion to our algorithm in Section \ref{trade-off}.  

\subsection{Feynman-Schrodinger Hybrid simulation}\label{hybrid_alg}

The Schrodinger approach requires exponential space due to the need to store the entire wave function. One can take advantage of limited entanglement to reduce the space required~\cite{chen201864, markov2018quantum}. This method is called the Feynman-Schrodinger hybrid algorithm.

If one divides the circuit into two sub-circuits and ignores the entangling gates connecting the sub-circuits, one can reduce the memory cost to $O(2^{n/2})$. How can we ``ignore" the entangling gates between the patches? For every such gate, one performs a Schmidt decomposition of the gate, e.g., decomposing the control-$Z$ gate as follows~\cite{chen201864,markov2018quantum}:
\begin{equation}
CZ=\ket{0}\bra{0}\otimes I+\ket{1}\bra{1}\otimes Z. \label{CZ_decomp}
\end{equation}
If there are $x$ entangling gates between the two patches, there will be $2^x$ patched circuit configurations. Now the space cost of running Schrodinger's approach on every patch becomes $O(2^{n/2})$. Afterwards, one could establish a Feynman path summation with $2^x$ terms from the results of the Schrodinger simulation on the patched circuits, where the number of operations required is now $2^x$. Therefore, the total number of operations becomes $2^{x+n/2}$, which is better than the direct simulation time scaling $2^n$ if the connectivity between the two patches is low enough such that $x<n/2$. Even though the number of operations required by the hybrid algorithm is greater than direct simulation if $x>n/2$, the memory cost is brought down to $2^{n/2}$. This is preferable for circuits with more than 50 qubits, even for the connectivity of the Google supremacy circuits~\cite{arute2019quantum}.   

Both of the two algorithms we discussed above together with general tensor-network algorithms~\cite{markov2008simulating,boixo2017simulation,chen2018classical,huang2020classical} all belong to a bigger class of simulation methods called the monotone methods~\cite{huang2020explicit}. Algorithms in this class only care about the number and locations of non-zero elements in the matrices that describe the gates in the circuit, not what values these elements take. This means they will not be sensitive to gate and circuit structures, e.g., Clifford gates. An unconditional lower bound on the time scaling for this class of method is obtained as $(n-2)(2^{n-3}-1)$~\cite{huang2020explicit}. Below we review another class of algorithms outside of monotone methods which works better for circuits dominated by Clifford circuits.

\subsection{Bravyi-Gosset algorithm}\label{BG_intro}

Clifford operations are unitaries generated by Hadamard, Phase and $CNOT$ gates. Clifford operations on $\ket{0}^{\otimes n}$ followed by measurement in the computational basis can be efficiently simulated classically~\cite{gottesman1998heisenberg}. The sets of states obtained by Clifford operations acting on the computational-basis states are called stabilizer states. Clifford operations plus one non-Clifford gate can boost Clifford circuits into universal quantum computation. One such gate is the $T$ gate, which is defined as:
\begin{equation}
T=\begin{pmatrix}
1 & 0\\
0 & e^{i\pi/4} 
\end{pmatrix}.
\end{equation}
Therefore, Clifford+$T$ circuits cannot be efficiently simulated unless $P=BQP$, where $BQP$ stands for bounded-error Quantum Polynomial Time. An algorithm for simulating Clifford+$T$ circuits with cost scaling exponentially with the number of $T$ gates but polynomially with the number of qubits is the Bravyi-Gosset algorithm~\cite{bravyi2016improved}. The algorithm is based on the stabilizer rank~\cite{bravyi2016trading, bravyi2016improved, bravyi2019simulation, kocia2020improved} of resource states $\ket{A}^{\otimes t}=(T\ket{0})^{\otimes t}$. The $T$ gates are implemented by magic state injection~\cite{bravyi2005universal}. The strong simulation version of the Bravyi-Gosset algorithm has time cost that scales linearly with the stabilizer rank of $\ket{A}^{\otimes t}$. This rank is upper bounded by $2^{0.47t}$.

The original Bravyi-Gosset algorithm can only tackle non-Clifford gates that are in the third level of the Clifford hierarchy~\cite{zhou2000methodology}, because only these gates can be implemented by injecting magic states. Alternatively, one can use the method based on Sum-over-Cliffords for unitaries~\cite{bravyi2019simulation}:
\begin{equation}
U=\sum_{j}a_jK_j.    
\end{equation}
where $K_j$'s are all Clifford gates and $a_j$'s are some complex coefficients. This decomposition of unitaries is able to deal with unitaries that are beyond the third level of the Clifford hierarchy.

The Bravyi-Gosset algorithm has the property that adding one non-Clifford 
gate increases the number of stabilizer states in the expansion of the wavefunction. On average, one $T$ gate multiplies the number of stabilizer states one needs to keep track of by a factor of $2^{0.47}$. 

The way to avoid branching is to insert stabilizer projectors. No matter how many stabilizer states there are in the decomposition from the previous step, a sum of $\kappa$ stabilizer projectors will always turn them into a sum of $\kappa$ stabilizer states. In Section \ref{alg} we introduce an algorithm that combines the stabilizer projector insertion with the recursion Feynman path-integral approach. By replacing the non-Clifford gates with stabilizer projectors, we incorporate the advantage of stabilizer-based simulation into Feynman path-integral simulation.

Unentangled quantum states have a polynomial space description, so does Clifford operations. The Gottesman-Knill theorem allows classical computers to simulate quantum systems with unlimited entanglement. However, in a sense, Clifford operation doesn't unleash the true power of entanglement and can be interpreted in a locally causal description~\cite{bell1964einstein,cuffaro2017significance}. In fact, any sets of operation that can be classically simulated have a locally causal description~\cite{cuffaro2017significance}. What's more, $T$ gate is hard for stabilizer-based simulation but it doesn't cause troubles for monotone simulation methods, $CNOT$ is hard for monotone methods while being easy for stabilizer-based algorithms. Due to the difference of the descriptive formalism of quantum mechanics, operations can be easy in one formalism but hard in another. Might it be advantageous to have a formalism that combines the locally causal properties of these different classically simulatable descriptions? In this paper, we try to step into the realm of combining monotone methods and stabilizer-based descriptions. We obtained the polynomial-space SPIR algorithm, which improves the time scaling of the previous polynomial-space recursive Feynman path-integral simulation~\cite{aaronson2017complexity}.

\subsection{Quantum supremacy experiment}

The Google ``quantum supremacy" experiment~\cite{arute2019quantum} implemented 20 cycles of random gates chosen from a universal gate set~(see Appendix \ref{sup_gates}), where each cycle contains one round of single-qubit gates on every qubit and one round of two-qubit gates in fixed repeating configurations. A linear cross-entropy fidelity of $\sim 0.002$ is achieved after measurement. The set of single-qubit gates used contains two Clifford gates and one non-Clifford gate. The two-qubit gates are all non-Clifford gates. One restriction on the random selection of single-qubit gates is that the same gate does not act on the same qubit in two consecutive cycles. The non-Clifford depth of the circuit is 40.

For this random circuit, there are 1113 single-qubit gates and 430 two-qubit gates. Out of the 1113 single-qubit gates, one third of them are non-Clifford. This means there are roughly 800 non-Clifford gates in the circuit. This is beyond the reach of the Bravyi-Gosset or the Sum-over-Clifford methods.

Motivated by this fact, the algorithms we introduce in this paper can serve as an extension to the Bravyi-Gosset or the Sum-over-Clifford method for circuits with a larger fraction of non-Clifford gates. However, as we shall see, the large fraction of non-Clifford gates of the circuits in \cite{arute2019quantum} remains an obstacle for stabilizer-based methods. 

\section{Stabilizer Projector decomposition for unitaries} \label{unitary}

We denote the full set of $n$-qubit stabilizer states as STAB, which gives an overcomplete basis for $n$-qubit states. The stabilizer projector basis, $\{\ket{\phi_i}\bra{\phi_i}\}_{\phi_i\in {\rm STAB}}$, provides an overcomplete basis for $n$-qubit operators. This basis is used to decompose density matrices in \cite{howard2017application} and forms stabilizer pseudomixtures, where the coefficients are all real. They can also be used for decomposing a unitary matrix as follows:
\begin{equation}
U=\sum_{i}c_i\ket{\phi_i}\bra{\phi_i}.   \label{Proj_decmop}
\end{equation}
where the $c_i$'s are complex coefficients and the $\ket{\phi_i}$s are stabilizer states. 

We define the stabilizer projector rank of a unitary operator as follows:
\begin{equation}
\kappa(U)=\text{min}\{k: U=\sum_{i=1}^{k}c_i\ket{\phi_i}\bra{\phi_i}\}. \label{StabProj_rank}     
\end{equation}
Because the matrix rank of a single stabilizer projector is one while the rank of a unitary matrix is $2^n$, we at least need $2^n$ terms to decompose a unitary operator. This implies that the stabilizer projector rank of an $n$-qubit unitary is at least $2^n$. Therefore it is clear that this quantity is not a measure of non-Cliffordness of the gate, because $\kappa\neq 1$ for Clifford gates. The stabilizer projector decomposition provides a representation of an arbitrary unitary that enables the action of the operator on the stabilizer states to be computed with cost $O(n^3\kappa)$.

We give upper bounds for the stabilizer projector rank for the non-Clifford gates used in the Google quantum supremacy experiment~\cite{arute2019quantum} in Table~\ref{supremacy}, where the gates $\sqrt{W}$, $\text{fSim}(\pi/2,\pi/6)$ are defined in Appendix \ref{sup_gates}. These non-Clifford gates are not diagonal, therefore the stabilizer projector decomposition is non-trivial.

\begin{table}[]
    \centering
    \begin{tabular}{c|c|c}
         Gate  & Upper bounds & Probability\\
         \hline
         $\text{fSim}(\pi/2,\pi/6)$  & 4 & 4/9\\
         \hline
         $\text{fSim}(\pi/2,\pi/6)\cdot\sqrt{W_1}\sqrt{W_2}$ & 10 & 1/9\\
         \hline
         $\text{fSim}(\pi/2,\pi/6)\cdot\sqrt{W_1}$ & 12 & 4/9\\
         \hline
         $\sqrt{W_1}\sqrt{W_2}$ & 6 & NA
    \end{tabular}
    \caption{Upper bounds on stabilizer projector rank and probability of occurrence among all two-qubit gates for the non-Clifford gates in the Google supremacy experiment. See Appendix \ref{sup_decomp} for one of the decompositions achieving these upper bounds for each gate.}
    \label{supremacy}
\end{table}

\section{Algorithm}\label{alg}

In this Section we discuss two simulation algorithms based on the stabilizer projector decomposition for unitaries introduced in Section~\ref{unitary}. We call the first algorithm the stabilizer path integral recursion~(SPIR) method and the second algorithm the stabilizer projector contraction~(SPC) method. Our polynomial-space SPIR algorithm improves the upper bound of the Aaronson-Chen polynomial-space Feynman path-integral simulation from $O((2d)^n)$ to $O((2d_{nc})^n)$. Here we omitted the non-exponential factors in the scaling and the exponents are exactly $n$ for diagonal non-Clifford gates but $\alpha n$ for non-diagonal cases where $\alpha>1$ is a constant the depends on the stabilizer projector decomposition, see Table \ref{scaling} for the exact number.

\subsection{Polynomial-space SPIR method} \label{trade-off}

To evaluate an amplitude $\bra{x}U\ket{0^n}$, one can replace all the non-Clifford gates in the circuit for $U$ with the projector decomposition of eq.~(\ref{Proj_decmop}) such that they form layers of stabilizer projectors. The amplitude then is given as a sum of products of Clifford operators contracted by stabilizer states. Calculating the inner product between two $n$-qubit stabilizer states takes $O(n^3)$ time~\cite{aaronson2004improved, garcia2012efficient}. Therefore the exponential part of the scaling of this simulation comes from the number of terms in the stabilizer projector decomposition of each gate. 

This procedure is analogous to a path integral formulation in eq.~(\ref{path_sim}), except here the unitaries are all Clifford gates and the computational basis states are replaced by general stabilizer states. Therefore the same structure of repeated calculations is also present in this procedure. We can use the recursion implementation of \cite{aaronson2017complexity} as follows. In general, if we divide the circuit $U$ we want to simulate into 2 depth $\ceil{\frac{d_{nc}}{2}}$ sub-circuits $U_1$ and $U_2$ with a depth-1 layer of non-Clifford gates $U_c=\sum_{i}c_i\ket{\phi_i}\bra{\phi_i}$ in the middle, we can now calculate an amplitude. The Clifford gates in the circuit will be ignored because they can be absorbed into the stabilizer states in $O(n)$ time.
\begin{equation}
\begin{split}
\bra{x}U\ket{0^n}&=\bra{x}U_2U_{nc}U_1\ket{0^n}\\
&=\sum_{i}c_i\bra{x}U_2\ket{\phi_i}\bra{\phi_i}U_1\ket{0^n}. 
\label{two_ly}
\end{split}
\end{equation}

The stabilizer projector rank of $U_{nc}$, which acts on $n$ qubits, is $\kappa_n=2^k$, which we assume is the maximum among all non-Clifford layers without loss of generality. We keep splitting the circuit at the non-Clifford layers, obtaining the following recursion relation 
\begin{equation}
T(d_{nc})=2\kappa_n T(d_{nc}/2),    
\end{equation}
The base case, which only contains one layer of non-Clifford gates, takes $T(1)=O(n^32^k)$ operations because one needs to calculate all $2^k$ inner-products for one layer of stabilizer projectors. Therefore the total number of operations to calculate this amplitude will be 
\begin{equation}
T(d_{nc})=O((2d_{nc})^{k+1}n^3)=O(\kappa_n d_{nc}^{k+1}n^3).
\end{equation}

We summarize this algorithm as follows:
\begin{enumerate}
    \item Divide the circuit into layers of Clifford gates and non-Clifford gates.
    \item Find the middle~($[d_{nc}/2]$th) layer of non-Clifford gates, divide the whole circuit into two sub-circuits, and rewrite this non-Clifford layer into a sum of stabilizer projectors. 
    \item Recursively implement step 2 for each sub-circuit and for each stabilizer projector until one gets down to the base case. For the base case, the $U_1$ and $U_2$ in eq.~(\ref{two_ly}) become Clifford layers $C_1$ and $C_2$. Calculate the inner-products for all $2^k$ projectors and sum them up to obtain the value for $\bra{\phi_j}C_2U_{nc}C_1\ket{\phi_k}$.
    \item Return the value obtained in the base cases from step 3 and recursively return the values to the level above, until one gets back to the root level and obtains the value for the final amplitude.
\end{enumerate}

This recursive method, is in spirit the same as the original recursive Feynman path-integral simulation, therefore the space requirement is $O(n\log d_{nc})$, as we already discussed in Section \ref{Recur_Path}. However, we manage to improve the time cost from $O((2d)^n)$ to $O((2d_{nc})^n)$, which makes it the state-of-the-art polynomial-space simulation of general random quantum circuits as far as we know. Previous work in \cite{shi2017recursive} is also a recursive method with polynomial space where the expensive gates are the non-diagonal gates. Their method are more preferable for quantum Fourier transform circuits~\cite{coppersmith2002approximate, nielsen2002quantum} or Instantaneous Quantum Polynomial~(IQP) circuits~\cite{shepherd2009temporally, bremner2011classical} because the number of non-diagonal gates is $O(n)$. Meanwhile, in our method and the original recursive Feynman path-integral method, one can further reduce the time cost by compressing adjacent layers of diagonal gates into one layer, in which case the effective non-Clifford depth of the IQP circuit is only 1. Therefore, our algorithm can simulate IQP circuits in $O(2^n)$ time with polynomial space requirement.

In subsection \ref{Exptime} we discuss an algorithm requiring exponential space, which may be used to combine with the Feynman-Schrodinger hybrid algorithm or the Sum-over-Clifford algorithm.

\subsection{Exponential-space SPC method}\label{Exptime}

The SPIR algorithm in the previous section is a Feynman-path type algorithm that calculates a single amplitude at a time. One can also evolve the whole state altogether as in the Schrodinger method using the stabilizer projector decomposition.

One first partitions the circuit into alternating rounds of one layer of Clifford gates followed by one layer of non-Clifford gates. Then one runs the following exponential-space SPC algorithm:
\begin{enumerate}
\item Start with computational basis state $\ket{0}^{\otimes n}$ and use the Gottesman-Knill theorem to evolve through the first Clifford layer at the beginning of the circuit, obtaining a stabilizer state $\ket{\psi_1}=\ket{\phi}$. This step takes a $O(m_1n^2)$ operations where $m_1$ is the number of Clifford gates in the first layer.
\item Decompose the next layer of non-Clifford gates into a sum of stabilizer projectors as in eq.~(\ref{StabProj_rank}), and project the stabilizer state we obtained in step 1 onto these projectors. The new state after this layer of non-Clifford gates is described as a sum of stabilizer states $\ket{\psi_2}=\sum_{i}c_i\braket{\phi_i}{\phi} \ket{\phi_i}$. This step takes $O(\kappa_n n^3)$ operations.
\item Evolve each stabilizer state in the sum for $\ket{\psi_2}$ above through the next round of Clifford layer separately. Then project each one onto the next non-Clifford layer with stabilizer projector decomposition. This step take $O(\kappa_n^2 n^3)$ operations.
\item Iterate step 3 until the end of the circuit.
\end{enumerate}
The overall number of operations is therefore $O(m_1n^2+\kappa n^3+(d_{nc}-1)\kappa^2 n^3)=O(d_{nc}\kappa^2 n^3)$ because we iterate step 3 $O(d_{nc})$ times. The space complexity is $O(\kappa)$.

The limitation of the improvement of our algorithm is, as we discussed before, the stabilizer projector rank of each layer of gates is at least $2^n$. Therefore, one wants to put as many non-Clifford gates as possible in one layer for stabilizer projector decomposition, and to ensure the gates act non-trivially on as many qubits as possible. In this way, one can avoid devoting resources to representing the identity. Even though, we remark that our SPC algorithm doesn't provide speedup compared to direct simulation or the improved stabilizer frame simulation~\cite{garcia2014simulation}. We present this algorithm to shed lights on future directions that may improve it to be preferential. As we will discuss, the key idea of the Feynman-Schrodinger hybrid algorithm applies naturally to our SPC method and can be therefore taken advantage of to reduce the space and time requirement.

\subsection{Combination of SPC method with Feynman-Schrodinger hybrid algorithm algorithm}

As introduced in Section \ref{hybrid_alg}, we can partition the initial circuit we need to simulate into two patches with an approximately equal number of qubits. The criterion for the partition is to make the number of entangling gates, $x$, that connects the two patches as small as possible. Then we decompose these entangling gates across the two patches into a sum of separable operations as in eq.~(\ref{CZ_decomp}). Now we have $2^x$ pairs of circuits, each of which contains two separate sub-circuits of $\sim n/2$ qubits. Then for each of these sub-circuits, we use our SPC method as a subroutine. Then we sum the amplitudes computed for all $2^x$ pairs. The space requirement of this algorithm is $\kappa_{n/2}$. The time requirement will be $O(2^x d_{nc}\kappa_{n/2}^2 n^3)$. As we mentioned before in Section \ref{hybrid_alg}, the space complexity is greatly reduced from $\sim 2^n$ to $\sim 2^{n/2}$, while the time complexity is reduced if $x<2\log_2 \kappa_{n/2}$. This is why it is important that we choose two patches of relatively equal size with $x$ as small as possible.

\subsection{Combination of SPC method with Sum-over-Clifford algorithm}\label{combine_BG}

The SPC algorithm is complementary to the original Sum-over-Clifford algorithm when the number of non-Clifford gates in the circuit exceeds the number of qubits. From another point of view, we can use the Sum-over-Clifford method to improve the time scaling of the SPC algorithm using its advantage for circuits with a small number of non-Clifford gates.
In fact, in step 1 and 2 of the SPC procedure we described above, the number of terms in the stabilizer decomposition of $\psi_1$ and $\psi_2$ suddenly goes from 1 to $\kappa$, which is of the order of $2^n$. Therefore, we will replace replace step 1 and 2 in the SPC method section by evolving the Sum-over-Clifford algorithm through the circuit until the number of Cliffords in the decomposition exceeds the stabilizer projector rank $\kappa$ of the next non-Clifford layer.
In the case of $T$-gates, this allows us to process $2.13n$ gates because one $T$-gate only adds a factor of $2^{0.47}$ to the rank on average. After that, we use steps 3 and 4 of the SPC procedure to prevent the rank from growing too fast.

\begin{table}[]
    \centering
    \begin{tabular}{c|c|c}
         \backslashbox{Algorithm}{cost}  & Time cost & Memory cost\\
         \hline
         Direct & $m2^n$ & $2^n$\\
         \hline
         Feynman path & $4^m$ & $m+n$\\
         \hline
         Hybrid~\cite{chen201864, markov2018quantum} & $2^{n/2+x}$ & $2^{n/2+1}$\\
         \hline
         Recursive path integral~\cite{aaronson2017complexity} & $d^n$ & $n\log d$\\
         \hline
         Tensor contraction~\cite{boixo2017simulation,chen2018classical} & $2^{tw(G)}$ & $2^{cw}$\\
         \hline
         Stabilizer rank~\cite{bravyi2016trading,bravyi2016improved} & $n^32^{0.47t}$ & $2^{0.47t}$\\
         \hline
         Our SPIR method & $n^3(2d_{nc})^{k}$ & $n\log d_{nc}$\\
         \hline
         Our SPC method & $d_{nc}2^{2k}n^3$ & $2^k$ 
    \end{tabular}
    \caption{Simulation cost for different algorithms. Here, $m$ is the total number of gates, $x$ is the number of entangling gates across the patches, $d$ is the circuit depth, $t$ is the number of non-Clifford gates in the circuit, $2^k$ is the stabilizer projector rank for each layer of gates, where $k\sim n$, and $d_{nc}$ is the non-Clifford gate-depth. $tw(G)$ is the treewidth of the undirected graph corresponding to the circuit~\cite{markov2008simulating}. $cw$ is the contraction width corresponding to a certain contraction order of the tensor, which is defined as the size of the biggest clique formed along the contraction. Big-$O$ notation is implicit in the table.}
    \label{scaling}
\end{table}

\section{Comparison to Bravyi-Gosset algorithm for a family of circuits}\label{Comparison_BG}

In this Section we will compare the time cost of our SPIR method with the strong version of the Bravyi-Gosset algorithm for a specific family of circuits. 

Both of the algorithms have an $n^3$ factor in the time cost, therefore it is sufficient to only compare the exponential part. The Bravyi-Gosset algorithm has exponential scaling $2^{0.47 t}$, while our method has scaling $(2d_{nc})^{k}$.

We will compare our SPIR method to the Bravyi-Gosset algorithm for simulation of an ensemble of random circuits with alternating single-qubit and two-qubit rounds. Specifically we consider the family of circuits where the single qubit gate-set contains both Clifford gates and non-Clifford gates and the non-Clifford gates are chosen with probability $p$. 

\begin{figure}[h]
\centering\includegraphics[height=5.5cm]{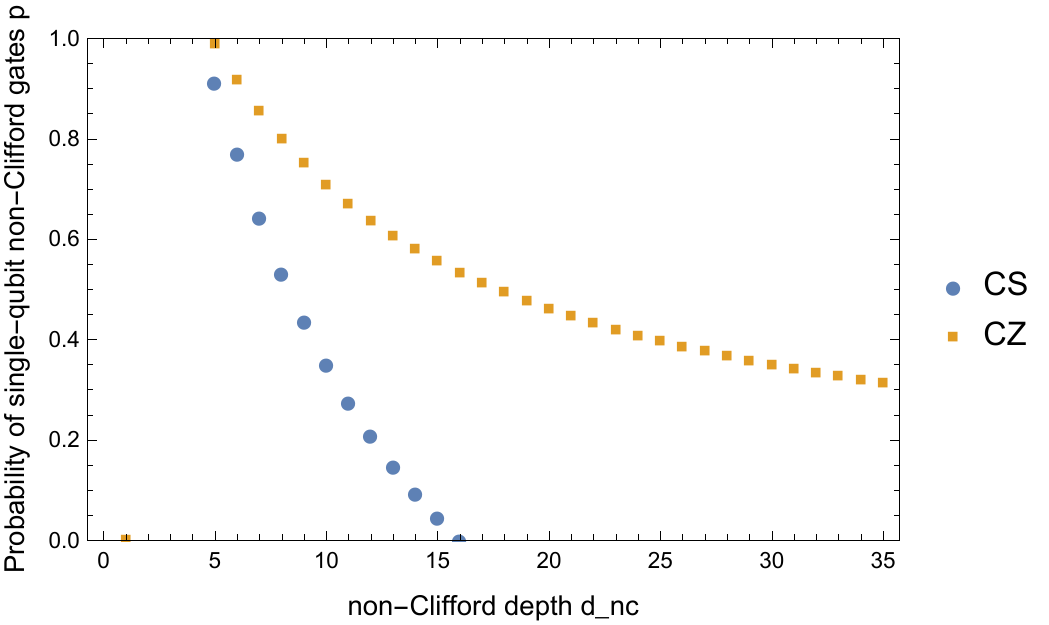}

\caption{In this plot we consider the ensemble random circuits with alternating rounds of single qubit and two qubit layers. We plot the threshold density~(probability) of non-Clifford gates for the SPIR method to have a better scaling than the strong Bravyi-Gosset algorithm for such circuits of different non-Clifford depth as in eq.~(\ref{CZ}) and eq.~(\ref{CS}). The two-qubit gates in comparison here are CZ and CS gates. Here the curve for CS is going negative for $d_{nc}>16$, which means for any density of single-qubit non-Clifford gates, the scaling of the SPIR method is better than the strong Bravyi-Gosset algorithm. For $p=1/3$ and CZ gates as two-qubit gates, the SPIR method starts to have a better scaling when the non-Clifford depth is bigger than 31. For $p=1/3$ and CS gates, the SPIR method starts to have a better scaling when the non-Clifford depth is bigger than 11. 
}
\label{Plot}
\end{figure}

We analyse the performance for an example in the family where two-qubit gates are Control-$Z$~($CZ$) gates and single qubit gates are $T$ gates with probability $p$ and Clifford gates with probability $1-p$. The only non-Clifford gates in the circuit are the $T$ gates. On average, for a circuit with $d_{nc}$ cycles, there are $d_{nc}np$ $T$ gates. We decompose the product of $T$ gates and $CZ$ gates in one cycle into a single unitary and perform a stabilizer projector decomposition. Therefore the effective depth in the SPIR method is $d_{nc}/2$. We also have $k=n$ because the product of $T$ gates and $CZ$ gates are all diagonal and can be represented exactly by the $2^n$ projectors of the computational basis states. Therefore the time complexities of our SPIR method and the Bravyi-Gosset algorithm are $(d_{nc})^{n}$ and $2^{0.47np d_{nc}}$ respectively. Hence the SPIR method will have a better scaling than the Bravyi-Gosset algorithm when 
\begin{equation}
\log_2 d_{nc}\leq 0.47 pd_{nc},  
\end{equation}
giving
\begin{equation}
p\geq \frac{\log_2 d_{nc}}{0.47 d_{nc}}. \label{CZ}
\end{equation}
This inequality is shown in Figure \ref{Plot}. Therefore if this inequality is satisfied, i.e., the density of non-Clifford gates is larger than this threshold with respect to the non-Clifford depth, the SPIR method has a better scaling, otherwise the Sum-over-Clifford method has the advantage.

Next we consider another example of the ensemble in which the two-qubit gates are Control-$S$~($CS$) gates, which is a non-Clifford gate. The number of non-Clifford gates in the circuits will be greatly increased and the non-Clifford depth is twice the number of cycles. In this case, for a $d_{nc}/2$ cycle circuit, there are on average $d_{nc}np/2$ $T$ gates and $nd_{nc}/4$ $CS$ gates. Meanwhile, due to the fact that the product of $T$ gates and $CS$ gates are all diagonal, we have $k=n$ for our algorithm. Therefore the time complexity of our algorithm and the Bravyi-Gosset algorithm are $(d_{nc})^{n}$ and $2^{0.47np d_{nc}/2+nd_{nc}/4}$ respectively. Hence our algorithm will have a better scaling when 
\begin{equation}
\log d_{nc}\leq 0.47 pd_{nc}/2+d_{nc}/4.   
\end{equation}
Giving
\begin{equation}
p\geq \frac{2(\log d_{nc}-d_{nc}/4)}{0.47 d_{nc}}. \label{CS}
\end{equation}
This inequality is also shown in Figure \ref{Plot}.

The comparison between the SPC method and Bravyi-Gosset algorithm is straightforward: it is the comparison between $\kappa_n^2$ and the stabilizer rank. Now we assume $T$ gates as the only non-Clifford resource. In this case, $\kappa_n=2^n$ for one layer of $T$ gates no matter how many qubits it acts on non-trivially. Meanwhile the stabilizer projectors in this decomposition are computational basis projectors $\ket{x}\bra{x}$. In the Bravyi-Gosset algorithm, one $T$ gate contributes a factor of $2^{0.47}$ to the time scaling. Hence, our algorithm performs better when $2^{2n}>2^{0.47t}$, i.e, when the number of $T$ gates is bigger than $4.26n$. If this is the case, we introduce the SPC method to complements the Sum-over-Clifford method, as we discussed in Section \ref{combine_BG}.

\section{Quantum supremacy experiment}

The random circuit in the Google supremacy experiment is one example in the ensemble of circuits we defined in Section \ref{Comparison_BG}. The two qubit gates are all non-Clifford gates. The one qubit gates are randomly selected from two Clifford gates and one non-Clifford gate. If we choose from these three gates with equal probability, we have $p=1/3$. The non-Clifford depth of the supremacy circuit is 20 if we combine the single-qubit gates and two-qubit gates in one cycle and decompose them into stabilizer projectors altogether as in Table I. 

Now we explicitly analyse the performance of our SPIR method for the supremacy circuits. In the average-case scenario, there will be roughly 10 $\text{fSim}(\pi/2,\pi/6)\cdot\sqrt{W_1}\cdot C_1$ gates~(Or $\text{fSim}(\pi/2,\pi/6)\cdot C_2\cdot\sqrt{W_2}$ gates), 2 $\text{fSim}(\pi/2,\pi/6)\cdot\sqrt{W_1}\sqrt{W_2}$ gates and 10 $\text{fSim}(\pi/2,\pi/6)\cdot C$ and 3 individual $\sqrt{W}$ gates in one cycle, where $C$, $C_1$, $C_2$ represent some Cliffords such as tensor product of $I$, $\sqrt{X}$ and $\sqrt{Y}$. Therefore, average-case stabilizer projector rank for a cycle is 
\begin{equation}
\kappa_n=2^k=12^{10}4^{10}10^{2}\cdot 18=2^{66.67}\approx 2^{1.25n}.
\end{equation} 
Therefore, the total cost for our SPIR method is roughly $O(2^{1.25n\log d_{nc}}n^3)$ and $O(40^{1.25n}n^3)=O(2^{6.7n}n^3)$ given $d_{nc}=20$. This cost makes the SPIR algorithm impractical for $n=53$.

If we try to tackle the supremacy circuit by the Sum-over-Clifford method, we have a rank-2 Clifford decomposition for the $\text{fSim}(\pi/2,\pi/6)$ gate and a rank-4 decomposition for the $\sqrt{W}$ gate, which are shown in Appendix \ref{sup_gates}. As we discussed in the introduction, there are 430 $\text{fSim}(\pi/2,\pi/6)$ 
gates. This alone gives an overhead of $2^{430}$, which is bigger than the exponential cost of our SPIR method. This indicates that the supremacy circuits are in a regime where the SPIR method scales better than the exact Sum-over-Clifford method. However, it is still impractical to simulate the supremacy circuits for both methods. 

The original recursive Feynman path-integral approach in \cite{aaronson2017complexity} gave a coefficient on the exponent for the supremacy circuit of $\log_2{2d}=\log_2(80)=6.32$, where $d$ is the total depth instead of the non-Clifford-gate depth. One can see our modified method is slightly worse than that of \cite{aaronson2017complexity}. This is because the non-Clifford-gate depth to total gate depth ratio is one half for the supremacy circuit, which is relatively large. 

For our SPC method, the time cost is $O(d_{nc}n^3\kappa_n^2)=O(d_{nc}n^32^{2.5n})$, from the discussion in the previous section. By combining the SPC method with the Feynman-Schrodinger algorithm, we obtain two equal-size sub-circuits with 35 cross-patch entangling gates~(see Figure S21 of the supplemental information of \cite{arute2019quantum}). Therefore the overall time scaling is $O(d_{nc}n^32^{35+1.25\times 53})\sim O(d_{nc}n^32^{100})$.

By combing the SPC method with Sum-over-Clifford, we can save one cycle of computation time. Because one cycle gives the exact Sum-over-Clifford method an overhead of $2^22\cdot 4^{18}=2^{58}$, two such cycles gives a factor of $2^{114}\sim 2^{2.5\times 53}$, which is the exponential overhead for one cycle in our SPC method. Again, as we remarked before, because the non-Clifford gates are relatively dense in the supremacy circuit, the benefit of combining the Sum-over-Clifford method with the SPC is limited. These two methods are prohibitive in the supremacy regime because the space requirement also scales as $O(2^{2.5n})$. This motivates for further research of approximate simulation with the SPC method which suits weak simulation.

\section{Discussion}

In this paper, we take advantage of Clifford gates for Feynman-path type simulators. We introduced the stabilizer projector rank of a unitary matrix and described two algorithms called SPIR and SPC algorithm based on that. By inserting stabilizer projector decompositions into the circuit, we keep the stabilizer rank from growing too rapidly even when the number of non-Clifford gates is larger than the number of qubits. The SPC algorithm, in this regard, complements the Bravyi-Gosset algorithm for circuits where the non-Clifford gates are more dense. However, this method also requires exponential space. In the spirit of the recursive Feynman-path algorithm proposed by Aaronson and Chen, the SPIR algorithm only uses polynomial space, but with super-exponential time cost. Although both our SPIR method and the SPC method have a larger scaling than the tensor network type simulation and Feynman-Schrodinger hybrid simulation for the supremacy circuit, we hope our algorithms offer some insights in regard of taking advantage of Clifford gates in Feynman-path type simulation, or in monotone methods in general. We expect to approach the unconditional bound for monotone methods, maybe even go beyond it with further developments along these ideas.

We hope our method of decomposing unitaries into stabilizer projectors will serve as a possible way to simulate NISQ devices. There is no reason to attempt to simulate hard NISQ devices like the Google circuits assuming the states and operations are perfect. One would want to take advantage of their low fidelity outputs. We made comparisons to the strong version of Bravyi-Gosset algorithm and Sum-over-Clifford algorithm in the paper. It is worth mentioning that the prominent part of the Sum-over-Clifford method is its performance for approximate simulation with weak simulation. Therefore in order to make a real comparison, a total-variational distance approximation version of our algorithm is worth exploring. By approximating the output probability distribution within a total variational distance or an additive error, we hope to have a algorithm that is comparable to the Sum-of-Clifford method for QAOA circuits. Meanwhile, being able to truncate and get rid of the smaller amplitudes in order to achieve some certain error threshold will enable us to compare our method with some of the state-of-the-art Matrix Product State~(MPS) simulations~\cite{guo2019general,zhou2020limits}. We leave these developments for future work.

\section*{Acknowledgements}

This work was supported by the National Science Foundation award number PHY 1720395. Y.H. thanks Will Kirby and Lucas Kocia for making helpful comments.

\bibliography{main}

\begin{thebibliography}{10}

\bibitem{aaronson2011computational}
Scott Aaronson and Alex Arkhipov.
\newblock The computational complexity of linear optics.
\newblock In {\em Proceedings of the forty-third annual ACM symposium on Theory
  of computing}, pages 333--342, 2011.

\bibitem{aaronson2017complexity}
Scott Aaronson and Lijie Chen.
\newblock Complexity-theoretic foundations of quantum supremacy experiments.
\newblock In {\em 32nd Computational Complexity Conference (CCC 2017)}. Schloss
  Dagstuhl-Leibniz-Zentrum fuer Informatik, 2017.

\bibitem{aaronson2004improved}
Scott Aaronson and Daniel Gottesman.
\newblock Improved simulation of stabilizer circuits.
\newblock {\em Physical Review A}, 70(5):052328, 2004.

\bibitem{arute2019quantum}
Frank Arute, Kunal Arya, Ryan Babbush, Dave Bacon, Joseph~C Bardin, Rami
  Barends, Rupak Biswas, Sergio Boixo, Fernando~GSL Brandao, David~A Buell,
  et~al.
\newblock Quantum supremacy using a programmable superconducting processor.
\newblock {\em Nature}, 574(7779):505--510, 2019.

\bibitem{barak2020spoofing}
Boaz Barak, Chi-Ning Chou, and Xun Gao.
\newblock Spoofing linear cross-entropy benchmarking in shallow quantum
  circuits.
\newblock {\em arXiv preprint arXiv:2005.02421}, 2020.

\bibitem{bell1964einstein}
John~S Bell.
\newblock On the einstein podolsky rosen paradox.
\newblock {\em Physics Physique Fizika}, 1(3):195, 1964.

\bibitem{boixo2018characterizing}
Sergio Boixo, Sergei~V Isakov, Vadim~N Smelyanskiy, Ryan Babbush, Nan Ding,
  Zhang Jiang, Michael~J Bremner, John~M Martinis, and Hartmut Neven.
\newblock Characterizing quantum supremacy in near-term devices.
\newblock {\em Nature Physics}, 14(6):595--600, 2018.

\bibitem{boixo2017simulation}
Sergio Boixo, Sergei~V Isakov, Vadim~N Smelyanskiy, and Hartmut Neven.
\newblock Simulation of low-depth quantum circuits as complex undirected
  graphical models.
\newblock {\em arXiv preprint arXiv:1712.05384}, 2017.

\bibitem{bravyi2019simulation}
Sergey Bravyi, Dan Browne, Padraic Calpin, Earl Campbell, David Gosset, and
  Mark Howard.
\newblock Simulation of quantum circuits by low-rank stabilizer decompositions.
\newblock {\em Quantum}, 3:181, 2019.

\bibitem{bravyi2016improved}
Sergey Bravyi and David Gosset.
\newblock Improved classical simulation of quantum circuits dominated by
  clifford gates.
\newblock {\em Physical review letters}, 116(25):250501, 2016.

\bibitem{bravyi2005universal}
Sergey Bravyi and Alexei Kitaev.
\newblock Universal quantum computation with ideal clifford gates and noisy
  ancillas.
\newblock {\em Physical Review A}, 71(2):022316, 2005.

\bibitem{bravyi2016trading}
Sergey Bravyi, Graeme Smith, and John~A Smolin.
\newblock Trading classical and quantum computational resources.
\newblock {\em Physical Review X}, 6(2):021043, 2016.

\bibitem{bremner2011classical}
Michael~J Bremner, Richard Jozsa, and Dan~J Shepherd.
\newblock Classical simulation of commuting quantum computations implies
  collapse of the polynomial hierarchy.
\newblock {\em Proceedings of the Royal Society A: Mathematical, Physical and
  Engineering Sciences}, 467(2126):459--472, 2011.

\bibitem{chen2018classical}
Jianxin Chen, Fang Zhang, Cupjin Huang, Michael Newman, and Yaoyun Shi.
\newblock Classical simulation of intermediate-size quantum circuits.
\newblock {\em arXiv preprint arXiv:1805.01450}, 2018.

\bibitem{chen201864}
Zhao-Yun Chen, Qi~Zhou, Cheng Xue, Xia Yang, Guang-Can Guo, and Guo-Ping Guo.
\newblock 64-qubit quantum circuit simulation.
\newblock {\em Science Bulletin}, 63(15):964--971, 2018.

\bibitem{coppersmith2002approximate}
Don Coppersmith.
\newblock An approximate fourier transform useful in quantum factoring.
\newblock {\em arXiv preprint quant-ph/0201067}, 2002.

\bibitem{cuffaro2017significance}
Michael~E Cuffaro.
\newblock On the significance of the gottesman--knill theorem.
\newblock {\em British Journal for the Philosophy of Science}, 68(1):91--121,
  2017.

\bibitem{fuchs2014introduction}
Christopher~A Fuchs, N~David Mermin, and R{\"u}diger Schack.
\newblock An introduction to qbism with an application to the locality of
  quantum mechanics.
\newblock {\em American Journal of Physics}, 82(8):749--754, 2014.

\bibitem{garcia2014simulation}
Hector~J Garcia and Igor~L Markov.
\newblock Simulation of quantum circuits via stabilizer frames.
\newblock {\em IEEE Transactions on Computers}, 64(8):2323--2336, 2014.

\bibitem{garcia2012efficient}
Hector~J Garcia, Igor~L Markov, and Andrew~W Cross.
\newblock Efficient inner-product algorithm for stabilizer states.
\newblock {\em arXiv preprint arXiv:1210.6646}, 2012.

\bibitem{gottesman1998heisenberg}
Daniel Gottesman.
\newblock The heisenberg representation of quantum computers.
\newblock {\em arXiv preprint quant-ph/9807006}, 1998.

\bibitem{guo2019general}
Chu Guo, Yong Liu, Min Xiong, Shichuan Xue, Xiang Fu, Anqi Huang, Xiaogang
  Qiang, Ping Xu, Junhua Liu, Shenggen Zheng, et~al.
\newblock General-purpose quantum circuit simulator with projected
  entangled-pair states and the quantum supremacy frontier.
\newblock {\em Physical review letters}, 123(19):190501, 2019.

\bibitem{hangleiter2019sample}
Dominik Hangleiter, Martin Kliesch, Jens Eisert, and Christian Gogolin.
\newblock Sample complexity of device-independently certified ``quantum
  supremacy''.
\newblock {\em Physical review letters}, 122(21):210502, 2019.

\bibitem{heisenberg1959physics}
Werner Heisenberg and Brian Bond.
\newblock {\em Physics and philosophy: the revolution in modern science}.
\newblock Allen \& Unwin St. Leonards, Australia, 1959.

\bibitem{howard2017application}
Mark Howard and Earl Campbell.
\newblock Application of a resource theory for magic states to fault-tolerant
  quantum computing.
\newblock {\em Physical review letters}, 118(9):090501, 2017.

\bibitem{huang2020explicit}
Cupjin Huang, Michael Newman, and Mario Szegedy.
\newblock Explicit lower bounds on strong quantum simulation.
\newblock {\em IEEE Transactions on Information Theory}, 2020.

\bibitem{huang2020classical}
Cupjin Huang, Fang Zhang, Michael Newman, Junjie Cai, Xun Gao, Zhengxiong Tian,
  Junyin Wu, Haihong Xu, Huanjun Yu, Bo~Yuan, et~al.
\newblock Classical simulation of quantum supremacy circuits.
\newblock {\em arXiv preprint arXiv:2005.06787}, 2020.

\bibitem{kocia2020improved}
Lucas Kocia.
\newblock Improved strong simulation of universal quantum circuits.
\newblock {\em arXiv preprint arXiv:2012.11739}, 2020.

\bibitem{leifer2014quantum}
Matthew~Saul Leifer.
\newblock Is the quantum state real? an extended review of $\psi$-ontology
  theorems.
\newblock {\em arXiv preprint arXiv:1409.1570}, 2014.

\bibitem{markov2018quantum}
Igor~L Markov, Aneeqa Fatima, Sergei~V Isakov, and Sergio Boixo.
\newblock Quantum supremacy is both closer and farther than it appears.
\newblock {\em arXiv preprint arXiv:1807.10749}, 2018.

\bibitem{markov2008simulating}
Igor~L Markov and Yaoyun Shi.
\newblock Simulating quantum computation by contracting tensor networks.
\newblock {\em SIAM Journal on Computing}, 38(3):963--981, 2008.

\bibitem{nest2008classical}
M~Nest.
\newblock Classical simulation of quantum computation, the gottesman-knill
  theorem, and slightly beyond.
\newblock {\em arXiv preprint arXiv:0811.0898}, 2008.

\bibitem{nest2009simulating}
M~Nest.
\newblock Simulating quantum computers with probabilistic methods.
\newblock {\em arXiv preprint arXiv:0911.1624}, 2009.

\bibitem{nielsen2011quantum}
Michael~A Nielsen and Isaac~L Chuang.
\newblock Quantum computation and quantum information: 10th.
\newblock {\em New York, NY, USA: Cambridge University Press},
  1107002176:9781107002173, 2011.

\bibitem{preskill2012quantum}
John Preskill.
\newblock Quantum computing and the entanglement frontier.
\newblock {\em arXiv preprint arXiv:1203.5813}, 2012.

\bibitem{preskill2018quantum}
John Preskill.
\newblock Quantum computing in the nisq era and beyond.
\newblock {\em Quantum}, 2:79, 2018.

\bibitem{pusey2012reality}
Matthew~F Pusey, Jonathan Barrett, and Terry Rudolph.
\newblock On the reality of the quantum state.
\newblock {\em Nature Physics}, 8(6):475--478, 2012.

\bibitem{schwarz2013simulating}
Martin Schwarz and Maarten Van~den Nest.
\newblock Simulating quantum circuits with sparse output distributions.
\newblock {\em arXiv preprint arXiv:1310.6749}, 2013.

\bibitem{shepherd2009temporally}
Dan Shepherd and Michael~J Bremner.
\newblock Temporally unstructured quantum computation.
\newblock {\em Proceedings of the Royal Society A: Mathematical, Physical and
  Engineering Sciences}, 465(2105):1413--1439, 2009.

\bibitem{shi2017recursive}
Andrew Shi.
\newblock Recursive path-summing simulation of quantum computation.
\newblock {\em arXiv preprint arXiv:1710.09364}, 2017.

\bibitem{terhal2002adaptive}
Barbara~M Terhal and David~P DiVincenzo.
\newblock Adaptive quantum computation, constant depth quantum circuits and
  arthur-merlin games.
\newblock {\em arXiv preprint quant-ph/0205133}, 2002.

\bibitem{villalonga2020establishing}
Benjamin Villalonga, Dmitry Lyakh, Sergio Boixo, Hartmut Neven, Travis~S
  Humble, Rupak Biswas, Eleanor~G Rieffel, Alan Ho, and Salvatore Mandr{\`a}.
\newblock Establishing the quantum supremacy frontier with a 281 pflop/s
  simulation.
\newblock {\em Quantum Science and Technology}, 5(3):034003, 2020.

\bibitem{zhou2000methodology}
Xinlan Zhou, Debbie~W Leung, and Isaac~L Chuang.
\newblock Methodology for quantum logic gate construction.
\newblock {\em Physical Review A}, 62(5):052316, 2000.

\bibitem{zhou2020limits}
Yiqing Zhou, E~Miles Stoudenmire, and Xavier Waintal.
\newblock What limits the simulation of quantum computers?
\newblock {\em arXiv preprint arXiv:2002.07730}, 2020.

\end{thebibliography}

\appendix

\section{Notions of Weak simulation}\label{notion}

The task for a weak simulator is to output a set of samples according to the quantum circuit.
If one can do strong simulation, then one can achieve weak simulation. Meanwhile, it is known that exact strong simulation is $\#P$-hard~\cite{nest2008classical, aaronson2011computational}, while the class BQP, is believed to be smaller than $\#P$-hard. 

The distinction between the strong simulation and weak simulation is not completely understood. Therefore it is not known to what extent one need approximate the probability distribution such that it is just enough to for the weak simulator to generate samples that are indistinguishable from the samples of the quantum device. However, there are several ways to weaken the condition in eq.~(\ref{Multi_error}) for the purpose of weak simulation, which we discuss as follows. 

First, our approximation $\hat{p}(x)$ has to be 0 if the actual $p(x)=0$, according to the multiplicative approximation in eq.~(\ref{Multi_error}). This is not necessary for weak simulation. We assume the number of samples required for the weak simulator is not exponential. In this case it maybe enough to spoof a verification test by sampling from a probability distribution that approximates the real probabilities better on the high probability~(heavy) outputs but more poorly on the low probability outputs. Hence, our estimation $\hat{p}(x)$ for $p(x)=0$ does not need to be exactly 0 but only needs to be small enough. Therefore we have the following weaker condition compared to eq.~(\ref{Multi_error}):
\begin{equation}
\abs{p(x)-\hat{p}(x)}<\epsilon_1\sim O(\frac{1}{\text{exp}(n)}).\label{Exp_additive} 
\end{equation}
What's more, one may not even need to approximate all high probability outputs well, but to approximate a majority of them~(more than half) better. For example, a total variational distance approximation suffices in most cases:
\begin{equation}
\sum_{x}\abs{p(x)-\hat{p}(x)}<\epsilon'.    \label{total}
\end{equation}
If there are only polynomial number of samples output by the quantum device and the output probability distribution only has non-negligible support on polynomial number of strings, then by the Chernoff?Hoeffding bound, one can only obtain information about the probabilities with polynomial precision~\cite{nest2009simulating,schwarz2013simulating}. Therefore it is enough to have a weak simulator that approximates the probabilities to the following degree:
\begin{equation}
\abs{p(x)-\hat{p}(x)}<\epsilon_2\sim O(\frac{1}{\text{poly}(n)}).\label{Poly_additive}
\end{equation}
This condition is much weaker than that of eq.~(\ref{total}) and (\ref{Exp_additive}). For a quantum circuit that has anti-concentrated output probability distribution, this constraint alone is not enough for a faithful weak simulator~\cite{hangleiter2019sample}. For example, we can imagine a distribution where almost all probabilities are exponentially small. The approximation in eq.~(\ref{Poly_additive}) may give us a distribution that only has supports on a polynomial number of outputs, say $q(n)$. Assume we sample a set of outputs that has size $10q(n)$ according to this distribution, which is still polynomial, we will see many repetitions in the sampled outputs.
When we sample from the actual quantum circuit, we will almost see no repetition because the actual probability distribution is anti-concentrated. 

As we mentioned before, we care more about the high probability outputs for the purpose of weak simulation. Therefore if we know the shape of the probability distribution, one could determine $l$ threshold probability values $0\leq y_1\leq y_2\leq...\leq y_l\leq 1$ and it may be enough in most cases to specify what range $p(x)$ is in among these $y_i$s for most probability distributions. For example, for an anti-concentrated probability distribution, if may suffice to know whether the probability $p(x)$ is bigger or smaller than the median probability of $p(x)$s for all $x$s~\cite{aaronson2017complexity}.

One can see there are several different notions and different conditions for the approximation of $p(x)$ for weak simulation. For some weak conditions like eq.~(\ref{Poly_additive}), the accuracy of the weak simulator depends on the shape of the actual distribution itself.

\section{Conventions and definitions for stabilizer states}

We can for convenience label the single qubit stabilizer states as follows: $\ket{+}=\ket{x}$, $\ket{-}=\ket{\bar x}$, $\ket{+i}=\ket{y}$, $\ket{-i}=\ket{\bar y}$, $\ket{0}=\ket{z}$, $\ket{1}=\ket{\bar z}$. Given a pair of labels of basis $a$ and $b$ where $a,b\in\{x,y,z\}$ we can define the following states:
\begin{equation}
\begin{split}
\ket{\Phi^{\pm}_{ab}}&=\frac{1}{\sqrt{2}}\left(\ket{ab}\pm \ket{\bar a\bar b}\right)\\    
\ket{\Phi^{\pm i}_{ab}}&=\frac{1}{\sqrt{2}}\left(\ket{ab}\pm i\ket{\bar a\bar b}\right)\\    
\ket{\Psi^{\pm}_{ab}}&=\frac{1}{\sqrt{2}}\left(\ket{a\bar b}\pm \ket{\bar a b}\right)\\    
\ket{\Psi^{\pm i}_{ab}}&=\frac{1}{\sqrt{2}}\left(\ket{a\bar b}\pm i\ket{\bar a b}\right)\\    \end{split}    
\end{equation}

This is the notation for two qubit stabilizer states that we will use below. There are six orthogonal bases of four stabilizer states that are maximally entangled, therefore the notation we use here doesn't correspond one-to-one to the stabilizer states.

\section{Gates in the Google supremacy experiment and their Sum-over-Clifford decompositions}\label{sup_gates}

The single-qubit Clifford gates are:
\begin{equation}
    \sqrt{X}=\frac{1}{\sqrt{2}}\begin{pmatrix}
1 & -i\\
-i & 1 
\end{pmatrix}\label{sqrt_X}
\end{equation}
and
\begin{equation}
    \sqrt{Y}=\frac{1}{\sqrt{2}}\begin{pmatrix}
1 & -1\\
1 & 1
\end{pmatrix}.\label{sqrt_Y}
\end{equation}
The non-Clifford single-qubit gate is:
\begin{equation}
\begin{split}
    \sqrt{W}&=T^{\dagger}\sqrt{X}T\\
    &=\frac{1}{\sqrt{2}}
\begin{pmatrix}
1 & -\sqrt{i}\\
\sqrt{-i} & 1
\end{pmatrix}.
\end{split}
\end{equation}

One Sum-over-Clifford decomposition for the $T$ gate is
\begin{equation}
\begin{split}
&T=\begin{pmatrix}
1 & 0\\
0 & e^{i\pi/4}
\end{pmatrix}
=e^{i\pi/8}\begin{pmatrix}
e^{-i\pi/8} & 0\\
0 & e^{i\pi/8}
\end{pmatrix}
\\&=\frac{e^{i\pi/8}}{2\cos(\pi/8)}(\begin{pmatrix}
1 & 0\\
0 & 1
\end{pmatrix}
+e^{i\pi/4}\begin{pmatrix}
1 & 0\\
0 & -1
\end{pmatrix})
=\frac{e^{i\pi/8}}{2\cos(\pi/8)}(I+e^{i\pi/4}Z).
\end{split}
\end{equation}

Therefore we have an upper bound for the rank of the sum-over-Clifford for the $\sqrt{W}$ gate as 4.

The two qubit gate is
\begin{equation}
\text{fSim}(\pi/2,\pi/6)=
\begin{pmatrix}
1 & 0 & 0 & 0\\
0 & 0 & -i & 0\\
0 & -i & 0 & 0\\
0 & 0 & 0 & e^{-i\pi/6}
\end{pmatrix}.
\end{equation}
which is non-Clifford.

This gate has sum-over-Clifford rank 2 because 
\begin{equation}
\begin{pmatrix}
1 & 0 & 0 & 0\\
0 & 1 & 0 & 0\\
0 & 0 & 1 & 0\\
0 & 0 & 0 & e^{-i\pi/6}
\end{pmatrix}    
=\frac{e^{i\pi/12}}{2\cos(\pi/12)}(I+e^{i\pi/6}CZ).
\end{equation}
and 
\begin{equation}
\text{fSim}(\pi/2,\pi/6)=iSWAP^{\dagger}\cdot\text{diag}\{1,1,1,e^{i\pi/6}\}.    
\end{equation}

\section{Stabilizer projector decomposition for non-Clifford gates in the Google supremacy experiment}\label{sup_decomp}

Here we give one stabilizer projector decomposition for each gate that gives the upper bounds in Table \ref{supremacy}. Notice the coefficients are not unique. The optimal support of the stabilizer projectors may also not be unique for the bounds.

\begin{equation}
\begin{split}
\text{fSim}(\pi/2,\pi/6)&=
\ket{zz}\bra{zz}+e^{i\pi/6}\ket{\bar{z}\bar{z}}\bra{\bar{z}\bar{z}}\\&-i\ket{\Psi_{zz}^{+}}\bra{\Psi_{zz}^{+}}+i\ket{\Psi_{zz}^{-}}\bra{\Psi_{zz}^{-}}
\end{split}
\end{equation}

\begin{equation}
\begin{split}
\sqrt{W_1}\sqrt{W_2}&=\frac{1}{2}
\begin{pmatrix}
1 & -\sqrt{i} & -\sqrt{i} & i\\
\sqrt{-i} & 1 & -1 & -\sqrt{i}\\
\sqrt{-i} & -1 & 1 & -\sqrt{i} \\
-i & \sqrt{-i} & \sqrt{-i} & 1 ,
\end{pmatrix}
\\&=-\sqrt{2}i\ket{\Phi_{zz}^{-}}\bra{\Phi_{zz}^{-}}+\ket{\Psi_{zz}^{-}}\bra{\Psi_{zz}^{-}}\\&+\frac{i}{\sqrt{2}}\ket{\Phi_{zz}^{+i}}\bra{\Phi_{zz}^{+i}}+(1+\frac{i}{\sqrt{2}})\ket{\Phi_{zz}^{-i}}\bra{\Phi_{zz}^{-i}}\\&-\sqrt{2}i\ket{xx}\bra{xx}+\sqrt{2}i\ket{\bar{y}\bar{y}}\bra{\bar{y}\bar{y}}
\end{split}
\end{equation}

\begin{equation}
\begin{split}
&\text{fSim}(\pi/2,\pi/6)\cdot\sqrt{W_1}\sqrt{W_2}=
(0.134-0.5i)\ket{zz}\bra{zz}\\&+(1.954+1.171i)\ket{\Phi_{zz}^{+}}\bra{\Phi_{zz}^{+}}-(1.644+0.329i)\ket{\Psi_{zz}^{+}}\bra{\Psi_{zz}^{+}}\\&+2i\ket{\Psi_{zz}^{-}}\bra{\Psi_{zz}^{-}}
-(1.866+0.5i)\ket{\Phi_{zz}^{+i}}\bra{\Phi_{zz}^{+i}}
\\&+(0.156+0.966i)\ket{\bar{x}\bar{x}}\bra{\bar{x}\bar{x}}
\\&+(1.903+0.778i)\ket{yy}\bra{yy}\\&+(2.869+2.451i)\ket{\bar{y}\bar{y}}\bra{\bar{y}\bar{y}}
\\&-(1.673+1.862i)\ket{\Psi_{yy}^{+}}\bra{\Psi_{yy}^{+}}\\&-(0.966+1.673i)\ket{\Psi_{xx}^{+i}}\bra{\Psi_{xx}^{+i}}.
\end{split}
\end{equation}

\begin{equation}
\begin{split}
&\text{fSim}(\pi/2,\pi/6)\cdot\sqrt{W_1}=(1.866+0.5i)\ket{zz}\bra{zz}\\&+(0.413+0.646i)\ket{x\bar{z}}\bra{x\bar{z}}+(1.319+0.354i)\ket{y\bar{z}}\bra{y\bar{z}}\\&+(1.378+2.319i)\ket{\bar{z}\bar{x}}\bra{\bar{z}\bar{x}}+(0.354-1.319i)\ket{z\bar{y}}\bra{z\bar{y}}\\&-(1.378+2.319i)\ket{\Phi_{zz}^{-}}\bra{\Phi_{zz}^{-}}-(1.220+2.112i)\ket{\Psi_{zz}^{+}}\bra{\Psi_{zz}^{+}}\\&-(0.159+0.207i)\ket{\Psi_{zz}^{-}}\bra{\Psi_{zz}^{-}}+(-0.036+2.319i)\ket{\bar{x}x}\bra{\bar{x}x}\\&+(1.061+1.319i)\ket{\bar{y}\bar{y}}\bra{\bar{y}\bar{y}}\\&-(1.378+0.905i)\ket{\Psi_{yy}^{-}}\bra{\Psi_{yy}^{-}}\\&-(0.354+0.095i)\ket{\Psi_{xx}^{-i}}\bra{\Psi_{xx}^{-i}}.
\end{split}
\end{equation}

\begin{equation}
\begin{split}
&\sqrt{W_2}\cdot iSWAP\cdot CZ\cdot\sqrt{W_1}=2\ket{zz}\bra{zz}+2\sqrt{2}\ket{\bar{x}\bar{y}}\bra{\bar{x}\bar{y}}\\&-((\sqrt{2}+1)+(\sqrt{2}-1)i)\ket{\Psi_{zz}^{-i}}\bra{\Psi_{zz}^{-i}}\\&+((\sqrt{2}+1)-(3+\sqrt{2})i)\ket{\Psi_{zz}^{+i}}\bra{\Psi_{zz}^{+i}}\\&-2(1+(\sqrt{2}-1)i)\ket{\Psi_{yy}^{-}}\bra{\Psi_{yy}^{-}}\\&+2(1+\sqrt{2}i)\ket{\Psi_{zz}^{-}}\bra{\Psi_{zz}^{-}}+2(-1+i)\ket{\Psi_{zx}^{-i}}\bra{\Psi_{zx}^{-i}}\\&+\sqrt{2}(-1+i)\ket{\Phi_{xz}^{-i}}\bra{\Phi_{xz}^{-i}}\\&+((-2-\sqrt{2})+0.586i)\ket{\Psi_{xz}^{-i}}\bra{\Psi_{xz}^{-i}}\\&+2(1+(\sqrt{2}-1)i)\ket{\Psi_{xx}^{+i}}\bra{\Psi_{xx}^{+i}}.
\end{split}
\end{equation}

\section{Stabilizer projector rank for $\sqrt{W}\otimes\sqrt{W}$}

Our starting point is the spectral decomposition of $\sqrt{W}\otimes\sqrt{W}$. The spectrum of $\sqrt{W}\otimes\sqrt{W}$ has a doubly-degenerate eigenvalue $1$ eigenspace and two eigenvectors with eigenvalues $\pm i$. The degenerate eigenspace is spanned by stabilizer states $\ket{\Psi^-}$ and $\ket{\Phi^{+i}}$. the non-degenerate, non-stabilizer eigenvectors are:
\begin{equation}
\begin{split}
\ket{v_{+i}}=\frac{1}{2}(e^{i\pi/4},-1,-1,e^{-i \pi/4})\\ 
\ket{v_{-i}}=\frac{1}{2}(e^{-i\pi/4},1,1,e^{i \pi/4})\\ 
\end{split}
\end{equation}
We can write the Pauli operator expansion of their contribution to the spectral decomposition as follows:
\begin{equation}
i(\ketbra{v_{+i}}{v_{+i}} -  \ketbra{v_{-i}}{v_{-i}})=-\frac{i}{2\sqrt{2}}(IX+XI+IY+YI) 
\end{equation}
The spectral decomposition of this Pauli expansion will of course be a stabilizer projector decomposition. We group commuting Pauli operators and use the fact that these sums have zero eigenvalues which reduces the number of terms in their spectral decomposition:
\begin{equation}
\begin{split}
\frac{1}{2}(IX+XI) &= \ketbra{xx}{xx}-\ketbra{\bar{x}\bar{x}}{\bar{x}\bar{x}}\\
\frac{1}{2}(IY+YI) &= \ketbra{yy}{yy}-\ketbra{\bar{y}\bar{y}}{\bar{y}\bar{y}}.\\
\end{split}
\end{equation}
Hence the stabilizer projector rank of $\sqrt{W}\otimes\sqrt{W}$ is six and the decomposition is:
\begin{equation}
\begin{split}
\sqrt{W}\otimes\sqrt{W} &= \ketbra{\Psi_{zz}^+}{\Psi_{zz}^+}+\ketbra{\Phi_{zz}^{+i}}{\Phi_{zz}^{+i}}\\
&-\frac{i}{\sqrt{2}}\biggl(\ketbra{xx}{xx}-\ketbra{\bar{x}\bar{x}}{\bar{x}\bar{x}}\\
&+\ketbra{yy}{yy}-\ketbra{\bar{y}\bar{y}}{\bar{y}\bar{y}}\biggr)\\
\end{split}
\end{equation}

\end{document}